\documentclass[11pt]{article}

\usepackage{acl}

\usepackage{times}
\usepackage{latexsym}
\usepackage{booktabs}
\usepackage{multirow}
\usepackage[T1]{fontenc}

\usepackage[utf8]{inputenc}

\usepackage{microtype}

\usepackage{inconsolata}

\usepackage{graphicx}

\title{Robustness and Reasoning Fidelity of Large Language Models in Long-Context Code Question Answering}

\author{
Kishan Maharaj\thanks{Equal contribution.} \:
Nandakishore Menon\footnotemark[1] \:
Ashita Saxena\footnotemark[1]\:
Srikanth Tamilselvam \\
IBM Research \\
\texttt{\{kishanmaharaj,Nandakishore,ashitasaxena\}@ibm.com} \\
\texttt{srikanth.tamilselvam@in.ibm.com}
}

\begin{document}
\maketitle
\begin{abstract}

Large language models (LLMs) increasingly assist software engineering tasks that require reasoning over long code contexts, yet their robustness under varying input conditions remains unclear. We conduct a systematic study of long-context code question answering using controlled ablations that test sensitivity to answer format, distractors, and context scale. Extending LongCodeBench’s Python dataset with new COBOL and Java question–answer sets, we evaluate state-of-the-art models under three settings: (i) shuffled multiple-choice options, (ii) open-ended questions and (iii) needle-in-a-haystack contexts containing relevant and adversarially irrelevant information. Results show substantial performance drops in both shuffled multiple-choice options and open-ended questions and brittle behavior in the presence of irrelevant cues. Our findings highlight limitations of current long-context evaluations and provide a broader benchmark for assessing code reasoning in both legacy and modern systems.

\end{abstract}

\section{Introduction}
\label{sec:introduction}
Large language models have achieved impressive results on a range of code understanding and generation tasks, driving their adoption in software engineering workflows. As these models are increasingly expected to operate on real-world codebases, their ability to reason over long code contexts—often consisting of thousands of lines spread across multiple files—has become a critical requirement. Effective long-context understanding demands not only scalability but also robust attention to sparse, task-relevant information embedded within large volumes of irrelevant code.

Early evaluations of code-oriented LLMs focused primarily on short or moderately sized inputs, limiting insight into their behavior under realistic context lengths. During the course of this project, LongCodeBench \cite{rando2025longcodebench} introduced a benchmark specifically designed for long-context code question answering in Python. It demonstrated that nominal context capacity does not guarantee effective reasoning across entire inputs. LongCodeU further expands this evaluation landscape by benchmarking long-context language models on diverse long code understanding tasks, revealing persistent performance degradation as context length grows and highlighting difficulties in maintaining global program coherence \cite{li2025longcodeu}. However, these benchmarks primarily measure accuracy under fixed task formulations and do not systematically probe robustness to input variations or adversarial conditions.

LongCodeBench offers valuable insights, but it also raises several open questions. First, its focus on a single programming language limits insight into whether long-context reasoning generalizes across modern and legacy ecosystems. Second, the fixed multiple-choice format may obscure weaknesses by allowing models to exploit superficial cues such as lexical overlap or positional patterns. Third, the benchmark does not explicitly evaluate robustness to irrelevant information, an essential property when analyzing real-world codebases containing large volumes of unrelated logic. Finally, legacy languages such as COBOL, which underpin critical infrastructure yet differ substantially in structure and idioms from modern languages, remain largely unexamined.

\begin{table*}[t]
\centering
\small
\begin{tabular}{l l c c c c c}
\hline
\textbf{Model} & \textbf{Method} & \textbf{32k} & \textbf{64k} & \textbf{128k} & \textbf{256k} & \textbf{512k} \\
\hline
Mistral-Small-24B & With Options & 64.6\% & 71.1\% & - & - & - \\
Mistral-Small-24B & Without Options & 45.1\% & 36.8\% & - & - & - \\
Gemini-2.5-Flash & With Options & 74.3\% & 68.4\% & 72.8\% & 66.2\% & 68.1\% \\
Gemini-2.5-Flash & Without Options & 47.8\% & 51.3\% & 54.3\% & 49.2\% & 40.4\% \\
GPT-4o & With Options & 61.9\% & 72.4\% & 72.5\% & - & - \\
GPT-4o & Without Options & 45.1\% & 50.0\% & 41.7\% & - & - \\
Gemini-2.5-Pro & With Options & 77.0\% & 73.7\% & 76.1\% & 69.2\% & 76.6\% \\
Gemini-2.5-Pro & Without Options & 53.1\% & 46.1\% & 54.3\% & 53.8\% & 44.7\% \\
Llama-3.1-405B & With Options & 73.5\% & 76.3\% & 68.5\% & - & - \\
Llama-3.1-405B & Without Options & 41.6\% & 46.1\% & 32.6\% & - & - \\
Claude-4.5-Sonnet & With Options & 80.5\% & 73.7\% & 79.3\% & - & - \\
Claude-4.5-Sonnet & Without Options & 42.5\% & 43.4\% & 46.7\% & - & - \\
\hline
\end{tabular}
\caption{Accuracy of models on Python LongCodeBench across context lengths. ``-'' indicates context length exceeded model limits.}
\label{tab:python_results}
\end{table*}

\begin{figure*}[t]
    \centering
    \includegraphics[width=\textwidth]{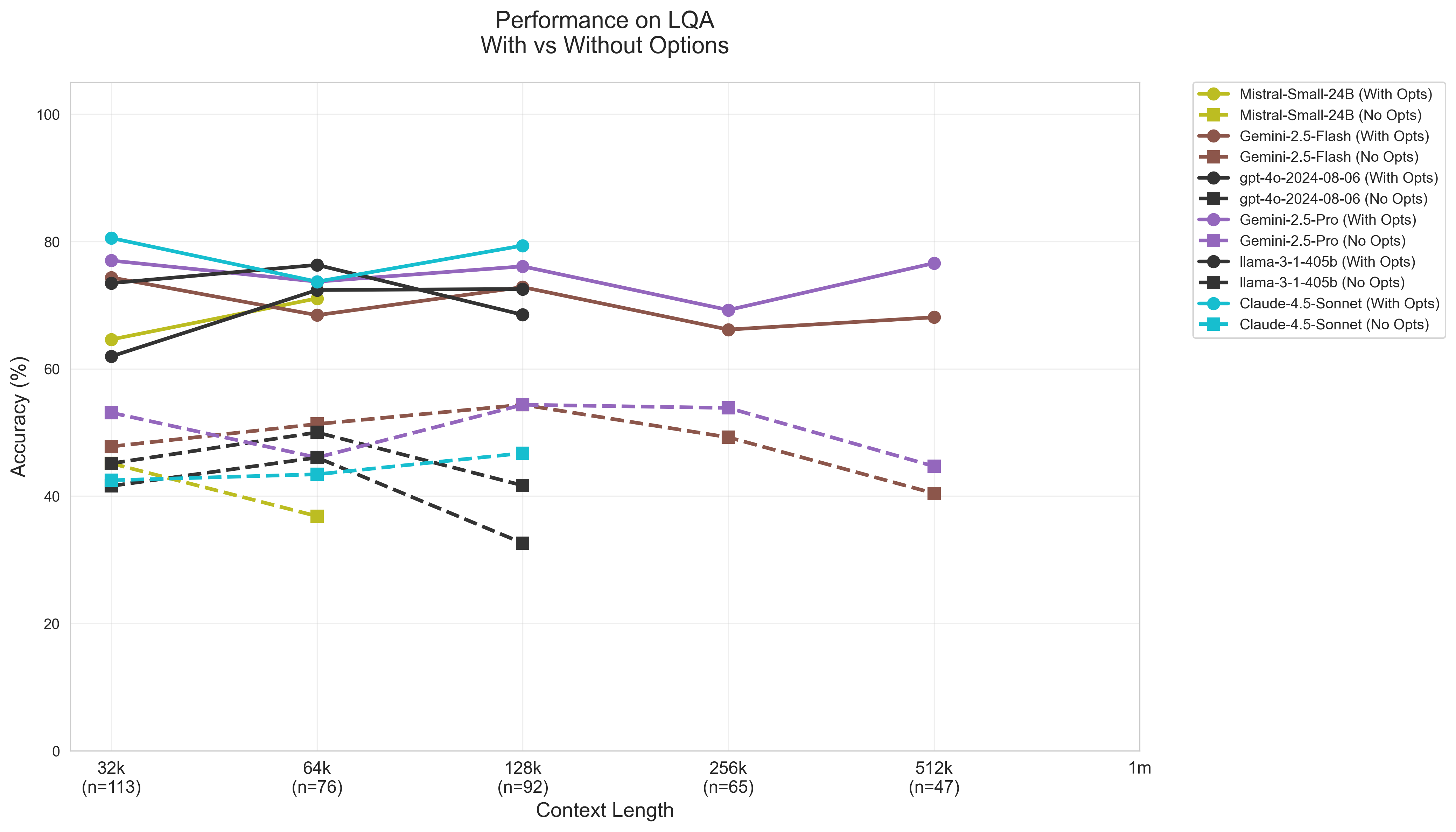}
    \caption{Accuracy trends across context lengths for Python dataset, comparing ``With Options'' and ``Without Options'' settings.}
    \label{fig:python_accuracy}
\end{figure*}

\begin{table*}[t]
\centering
\small
\begin{tabular}{l l c c c c c}
\hline
\textbf{Model} & \textbf{Method} & \textbf{32k} & \textbf{64k} & \textbf{128k} & \textbf{256k} & \textbf{512k} \\
\hline
Mistral-Small-24B & With Options & 95.1\% & 90.7\% & 95.7\% & - & - \\
Mistral-Small-24B & Without Options & 55.7\% & 50.0\% & 47.1\% & - & - \\
Gemini-2.5-Flash & With Options & 98.4\% & 92.6\% & 98.6\% & 91.1\% & 94.0\% \\
Gemini-2.5-Flash & Without Options & 67.2\% & 68.5\% & 54.3\% & 69.6\% & 38.0\% \\
GPT-4o & With Options & 96.7\% & 92.6\% & 87.1\% & - & - \\
GPT-4o & Without Options & 49.2\% & 57.4\% & 50.0\% & - & - \\
Gemini-2.5-Pro & With Options & 98.4\% & 92.6\% & 98.6\% & 95.5\% & 94.0\% \\
Gemini-2.5-Pro & Without Options & 62.3\% & 70.4\% & 57.1\% & 71.4\% & 32.0\% \\
\hline
\end{tabular}
\caption{Accuracy on OPPSCAL COBOL dataset across context lengths.}
\label{tab:oppscal_results}
\end{table*}

\begin{table*}[t]
\centering
\small
\begin{tabular}{l l c c c c c}
\hline
\textbf{Model} & \textbf{Method} & \textbf{32k} & \textbf{64k} & \textbf{128k} & \textbf{256k} & \textbf{512k} \\
\hline
Mistral-Small-24B & With Options & 73.1\% & 68.9\% & - & - & - \\
Mistral-Small-24B & Without Options & 46.2\% & 35.6\% & - & - & - \\
Gemini-2.5-Flash & With Options & 76.9\% & 77.8\% & 88.0\% & 85.7\% & 90.9\% \\
Gemini-2.5-Flash & Without Options & 50.0\% & 42.2\% & 38.7\% & 52.1\% & 50.3\% \\
GPT-4o & With Options & 80.8\% & 82.2\% & 85.3\% & - & - \\
GPT-4o & Without Options & 38.5\% & 51.1\% & 54.7\% & - & - \\
Gemini-2.5-Pro & With Options & 73.1\% & 77.8\% & 88.0\% & 90.8\% & 93.7\% \\
Gemini-2.5-Pro & Without Options & 50.0\% & 48.9\% & 50.7\% & 52.9\% & 52.6\% \\
Claude-4.5-Sonnet & With Options & 76.9\% & 80.0\% & 86.7\% & - & - \\
Claude-4.5-Sonnet & Without Options & 50.0\% & 46.7\% & 53.3\% & - & - \\
\hline
\end{tabular}
\caption{Accuracy on internal IBM COBOL dataset across context lengths.}
\label{tab:internal_results}
\end{table*}

\begin{figure*}[t]
    \centering
    \includegraphics[width=\textwidth]{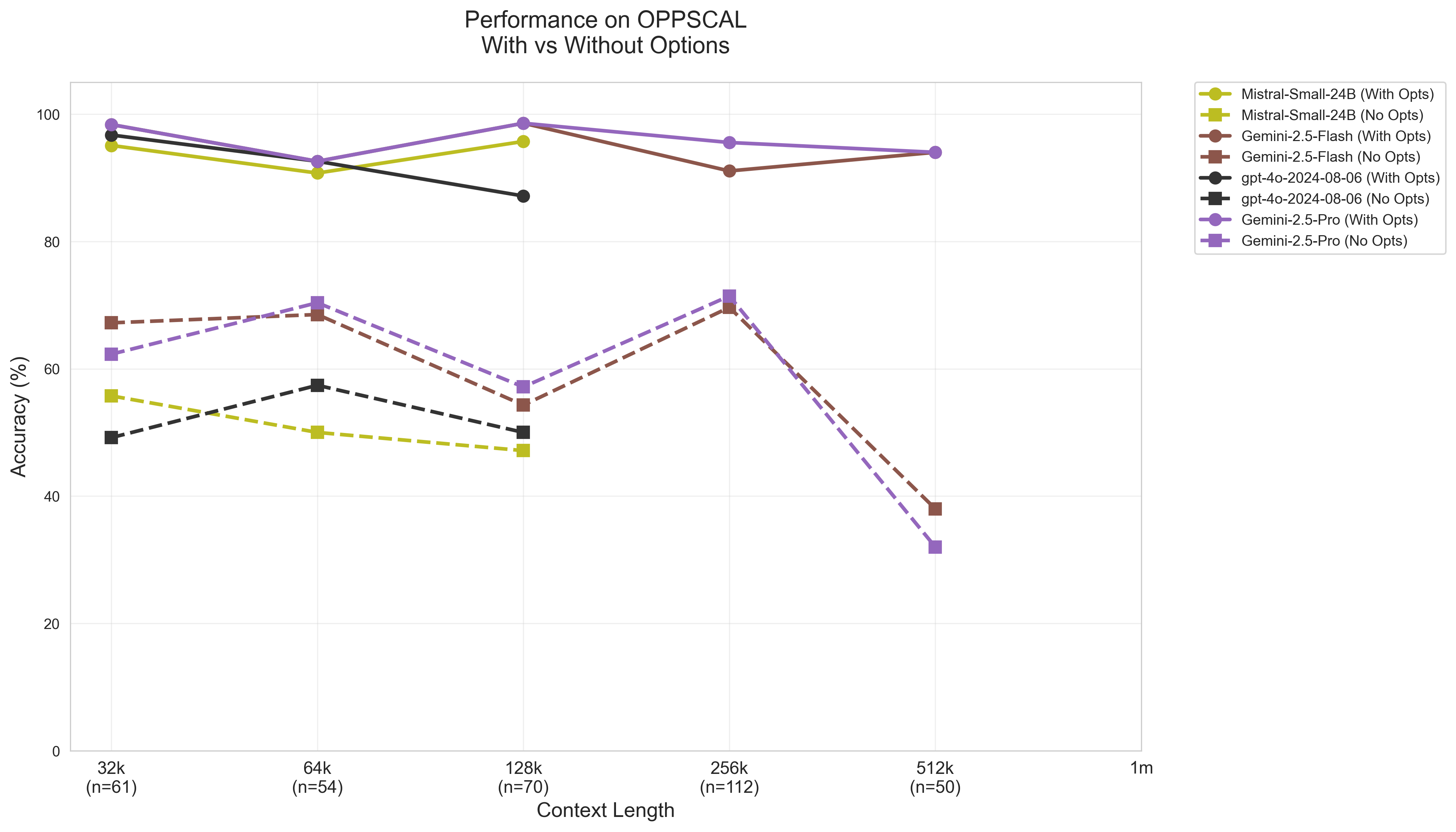}
    \caption{Accuracy trends for OPPSCAL dataset.}
    \label{fig:oppscal_accuracy}
\end{figure*}

\begin{figure*}[t]
    \centering
    \includegraphics[width=\textwidth]{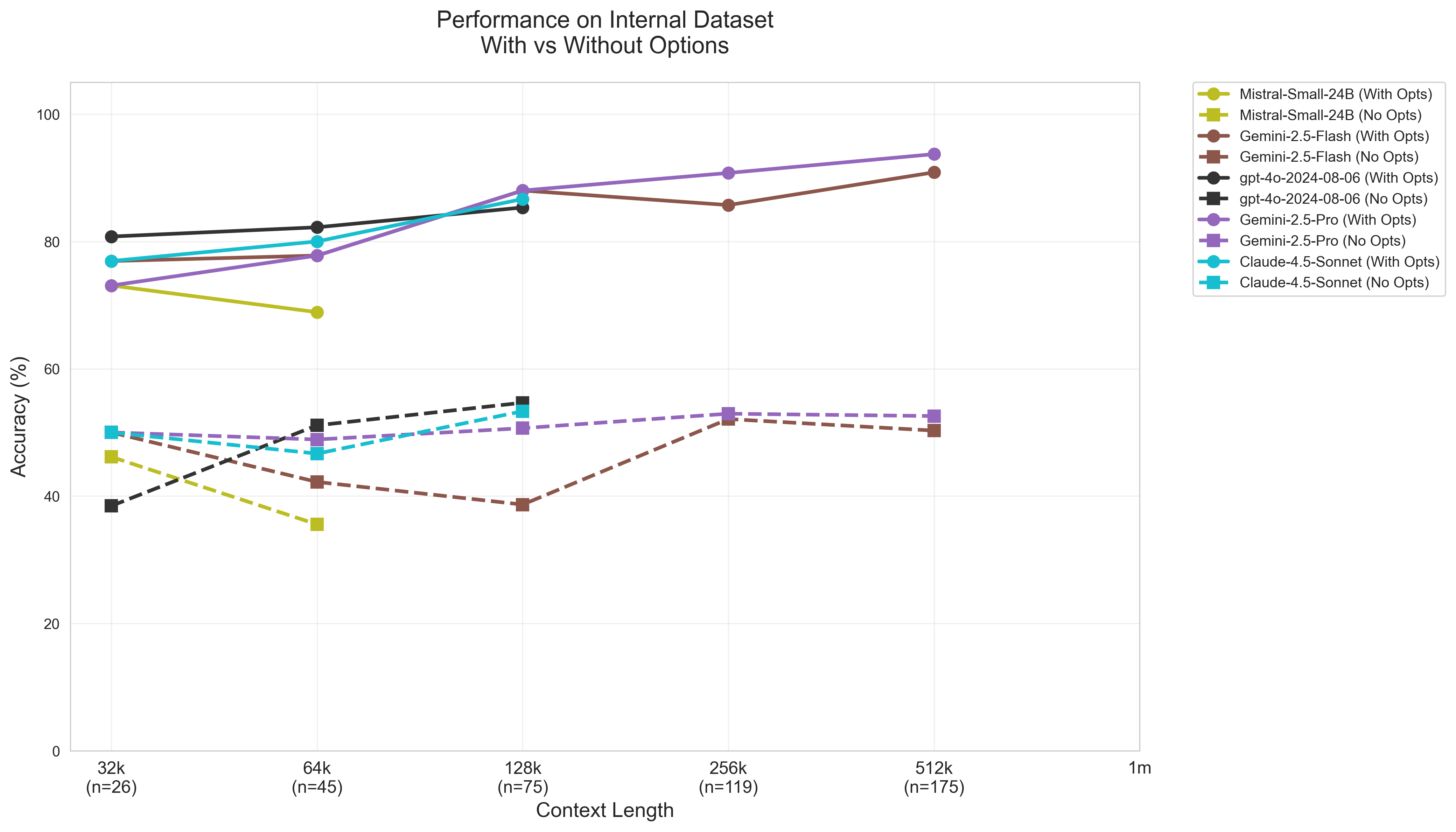}
    \caption{Accuracy trends for internal IBM COBOL dataset.}
    \label{fig:internal_accuracy}
\end{figure*}

\begin{figure*}[t]
    \centering
    \includegraphics[width=\textwidth]{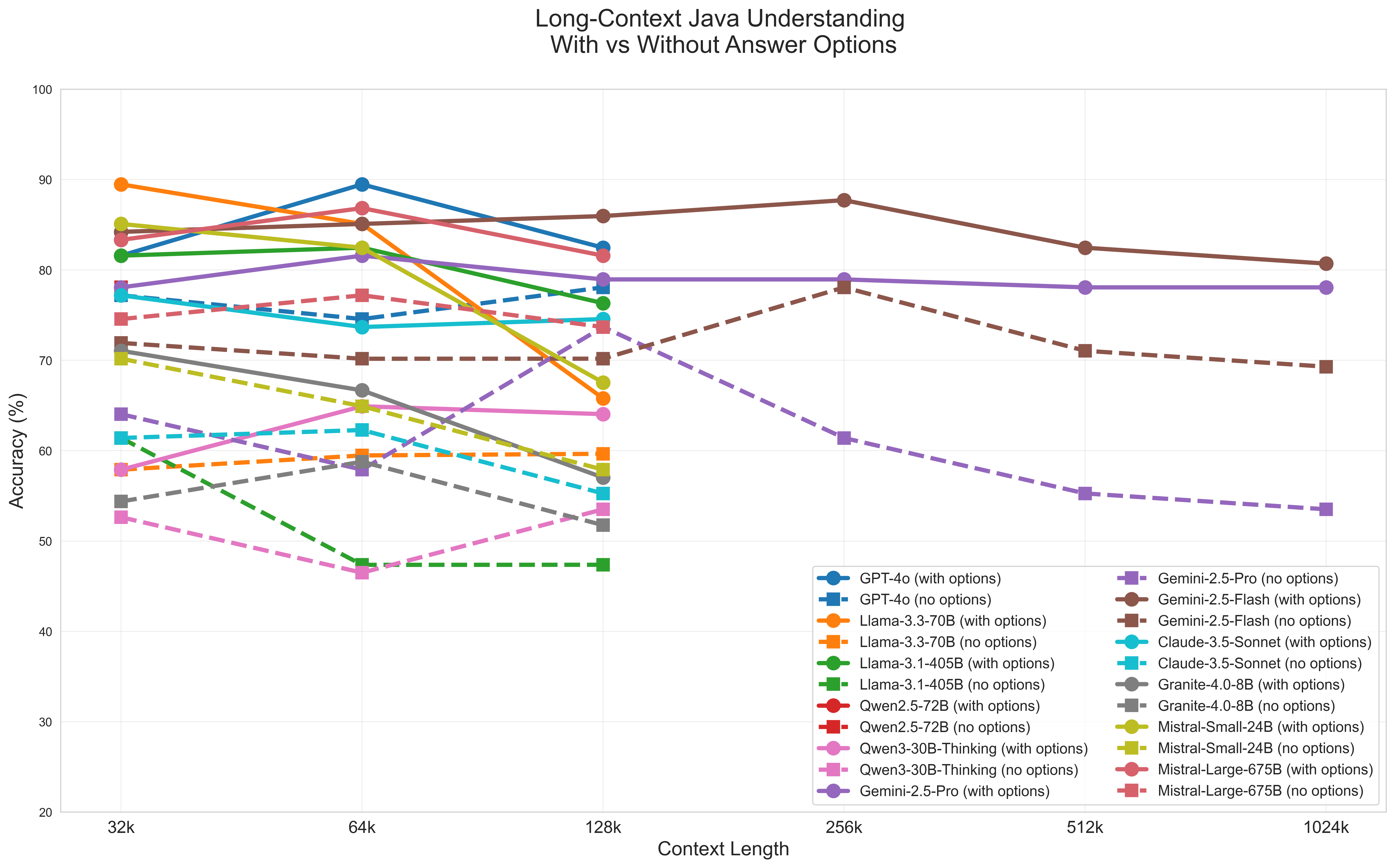}
    \caption{Accuracy trends across context lengths for all models on LongCodeBenchQA-Java. The figure highlights performance scaling from 32k to 1024k tokens for both multiple-choice and open-ended evaluation.}
    \label{fig:model-performance-wide}
\end{figure*}
\begin{table*}[t]
\centering
\small

\begin{tabular}{l l c c c c c c}
\hline
\textbf{Model} & \textbf{Method} & \textbf{32k} & \textbf{64k} & \textbf{128k} & \textbf{256k} & \textbf{512k} & \textbf{1024k} \\
\hline
Qwen2.5-72B & With Options & 83.3\% & - & - & - & - & - \\
Qwen2.5-72B & Without Options & 78.1\% & - & - & - & - & - \\
GPT-4o & With Options & 81.6\% & 89.5\% & 82.5\% & - & - & - \\
GPT-4o & Without Options & 77.2\% & 74.6\% & 78.1\% & - & - & - \\
Llama-3.3-70B & With Options & 89.5\% & 85.1\% & 65.8\% & - & - & - \\
Llama-3.3-70B & Without Options & 57.9\% & 59.5\% & 59.6\% & - & - & - \\
Llama-3.1-405B & With Options & 81.6\% & 82.5\% & 76.3\% & - & - & - \\
Llama-3.1-405B & Without Options & 61.4\% & 47.4\% & 47.4\% & - & - & - \\
Gemini-2.5-Pro & With Options & 78.1\% & 81.6\% & 79.0\% & 79.0\% & 78.1\% & 78.1\% \\
Gemini-2.5-Pro & Without Options & 64.0\% & 57.9\% & 73.7\% & 61.4\% & 55.3\% & 53.5\% \\
Gemini-2.5-Flash & With Options & 84.2\% & 85.1\% & 86.0\% & 87.7\% & 82.5\% & 80.7\% \\

Gemini-2.5-Flash & Without Options & 71.9\% & 70.2\% & 70.2\% & 78.1\% & 71.1\% & 69.3\% \\
Claude-4.5-Sonnet & With Options & 77.2\% & 73.7\% & 74.6\% & - & - & - \\

Claude-4.5-Sonnet & Without Options & 61.4\% & 62.3\% & 55.3\% & - & - & - \\
Qwen3-30B-Thinking & With Options & 57.9\% & 64.9\% & 64.0\% & - & - & - \\
Qwen3-30B-Thinking & Without Options & 52.6\% & 46.5\% & 53.5\% & - & - & - \\
Granite-4.0-8B & With Options & 71.0\% & 66.7\% & 57.0\% & - & - & - \\
Granite-4.0-8B & Without Options & 54.4\% & 58.8\% & 51.8\% & - & - & - \\
Mistral-Small-24B & With Options & 85.1\% & 82.5\% & 67.5\% & - & - & - \\
Mistral-Small-24B & Without Options & 70.2\% & 64.9\% & 57.9\% & - & - & - \\
Mistral-Large-675B & With Options & 83.3\% & 86.8\% & 81.6\% & - & - & - \\
Mistral-Large-675B & Without Options & 74.6\% & 77.2\% & 73.7\% & - & - & - \\
\hline
\end{tabular}
\caption{Java QA accuracy across context lengths for multiple-choice (With Options) and open-ended (Without Options) settings. Results span context windows from 32k to 1024k tokens. ``-'' indicates that the required context length exceeds the model's maximum input capacity.}

\label{tab:results}
\end{table*}

To address these gaps, we extend LongCodeBench along multiple dimensions. We retain the original Python dataset and introduce new long-context question-answer benchmarks for COBOL and Java derived from enterprise systems and large-scale open-source repositories. This multilingual expansion enables systematic evaluation of whether long-context reasoning capabilities transfer across programming paradigms and code styles.

We further introduce controlled perturbations to probe model behavior beyond raw accuracy. First, we shuffle multiple-choice answer options to test reliance on positional or lexical biases. Second, we remove answer options entirely, converting the task into open-ended question answering that requires explicit answer generation. Third, inspired by 'needle in a haystack' experiment \cite{needle-in-haystack}, we insert both relevant and adversarially irrelevant code fragments at varying depths within long contexts to assess distraction robustness.

We evaluate multiple state-of-the-art models across these settings and analyze performance as a function of context length, task formulation, language and evidence position. Across datasets, we observe three consistent failure modes. First, models exhibit a large recognition–generation gap: performance drops sharply when answer options are removed, indicating reliance on matching rather than reconstructive reasoning. Second, models demonstrate strong sensitivity to irrelevant information, often failing even when the correct evidence is present. Third, retrieval effectiveness is highly position-dependent, with pronounced recency bias and degraded performance for earlier context segments particularly in legacy languages such as COBOL.

We evaluate several state-of-the-art models across these settings, examining performance trends as a function of context length, task formulation, and needle placement. Our findings reveal that models often rely on fragile cues exposed by multiple-choice formats and exhibit pronounced degradation in the presence of irrelevant needles, even when the relevant information is present. These behaviors are consistent across languages and are particularly evident in COBOL, underscoring challenges in long-context generalization.

To summarize, our contributions are as follows :
\begin{itemize}
    \item \textbf{LongContextCodeQA}\footnote{\url{https://huggingface.co/datasets/mjkishan/LongContextCodeQA}} - A multilingual extension of LongCodeBench with new COBOL and Java long-context QA datasets drawn from real-world systems. 
    \item  A suite of controlled robustness evaluations, including option shuffling, open-ended generation, and distractor injection.
    \item A comprehensive empirical study showing consistent recognition–generation gaps, positional biases and vulnerability to irrelevant information across state-of-the-art models.
\end{itemize}



\section{LongCodeBench}
\label{sec:longcodebench}

\subsection{Dataset Overview}
LongCodeQA from LongCodeBench is a comprehensive Q\&A benchmark with four options designed to evaluate the performance of large language models on code understanding tasks requiring extended context windows up to 1 million tokens. Each question is paired with code contexts of varying lengths—32k, 64k, 128k, 256k, 512k and up to 1 million tokens—enabling systematic analysis of how model performance scales with context size.




\subsection{Experiment}
The original LongCodeBench evaluation protocol measures accuracy across different context length buckets using the multiple-choice format. Models are presented with a question and four possible answers, requiring them to identify the single correct option. The experiments reveal several key findings: (i) performance does not uniformly improve with larger context windows, suggesting challenges in effective information retrieval; (ii) models exhibit varying degrees of robustness to context length, with some showing significant degradation beyond certain thresholds; and (iii) architectural differences between model families lead to distinct scaling behaviors.

The benchmark enables comparison across state-of-the-art models including GPT-4o, Claude, Gemini, Llama, and other frontier systems, providing insights into their relative strengths in long-context code understanding.

\subsection{LongContextCodeQA}
While the original LongCodeBench provides valuable insights into long-context performance, it leaves open questions about the robustness and reasoning fidelity of model predictions. Specifically, the multiple-choice format may allow models to exploit superficial cues—such as lexical overlap between questions and answer options—rather than demonstrating genuine code comprehension.

To probe these concerns, we extend LongCodeBench by studying two critical variations:

\paragraph{With Options (Shuffled).} We randomise the order of answer choices to test sensitivity to positional bias and ordering effects. Robust models should maintain stable performance regardless of option arrangement. Significant performance changes under shuffling indicate reliance on non-semantic cues.

\paragraph{Without Options.} We remove the multiple-choice options entirely, converting the task into open-ended question answering. Models must generate answers directly from the code context rather than selecting from candidates. Correctness is evaluated using an LLM-as-a-Judge framework, where a separate model assesses whether the generated answer aligns with the ground truth. This variation exposes the \textit{recognition versus generation gap}—models that perform well on multiple-choice may struggle when required to synthesize answers independently, revealing reliance on recognition-based shortcuts rather than deep reasoning.

Together, these extensions provide a more realistic assessment of long-context code understanding, reflecting scenarios in which developers must interpret unfamiliar codebases without predefined answer choices.

\section{Datasets}
\label{sec:datasets}
To evaluate long-context reasoning across diverse programming paradigms, we construct new benchmarks for COBOL and Java and combine them with the Python dataset from LongCodeBench. The resulting suite spans modern open-source systems and legacy enterprise software, enabling systematic analysis of cross-language generalisation, domain effects, and context-scale robustness.

All datasets are designed to require multi-file reasoning over large code contexts, mirroring real-world scenarios in which relevant information is sparse and distributed.

\subsection{LongContextCodeQA: COBOL}
\label{sec:cobol_dataset}
Legacy COBOL systems remain critical to sectors such as finance, healthcare and government yet they differ substantially from modern languages in syntax, structure and programming idioms. To evaluate model performance on this underrepresented domain, we construct two datasets from production-grade COBOL codebases.

\paragraph{OPPSCAL Dataset.} The OPPSCAL dataset is derived from the Medicare Pricer codebase used by the Centers for Medicare \& Medicaid Services (CMS). The source code is publicly available from CMS~\footnote{\url{https://www.cms.gov/files/zip/4th-quarter-opps-2021-files-10202021.zip}} and contains over 512,000 tokens of COBOL code implementing complex pricing algorithms for hospital outpatient services. We manually curated multiple-choice questions targeting record structures, data flow logic and computation procedures specific to healthcare pricing systems. The questions require models to trace variable assignments across multiple divisions, understand COBOL-specific constructs (e.g., PERFORM loops, REDEFINES clauses), and reason about conditional logic embedded in procedural code.

Context lengths for this dataset span 32k, 64k, 128k, 256k, and 512k tokens, with questions distributed across these buckets to assess scaling behavior. The dataset contains questions across all buckets, providing comprehensive coverage of long-context scenarios.

\paragraph{Internal Enterprise Dataset.} To complement the public dataset, we construct a second benchmark from an internal enterprise COBOL system used for high-volume transaction processing. The questions are designed to test understanding of CICS (Customer Information Control System) transaction processing, database interactions and multi-program call patterns common in enterprise COBOL applications.

Both COBOL datasets follow the same experimental protocol as the Python dataset: each question is evaluated in two settings— multiple-choice, (results averaged over multiple tests with shuffled options), and without options. This enables direct comparison of model robustness across modern and legacy programming languages.

\subsection{LongContextCodeQA: Java}
\label{sec:java_dataset}

To evaluate performance on contemporary software ecosystems, we construct a Java dataset from four of the most-starred public Java repositories on GitHub: Elasticsearch\footnote{\url{https://github.com/elastic/elasticsearch}} (75.6k stars, Apache 2.0 license), Cassandra\footnote{\url{https://github.com/apache/cassandra}} (19.5k stars, Apache 2.0 license), Dubbo\footnote{\url{https://github.com/apache/dubbo}} (41.6k stars, Apache 2.0 license), and Kafka\footnote{\url{https://github.com/apache/kafka}} (31.4k stars, Apache 2.0 license).

From these repositories, we curated 114 questions covering diverse aspects of Java software engineering: code understanding and API behavior (37.7\%), exception handling (19.3\%), javadoc annotations (12.3\%), packaging and deployment (12.3\%), Java version-specific features (7.9\%), testability and design patterns (6.1\%), and configuration design (4.4\%).

Each question is paired with code contexts truncated to six target lengths: 32k, 64k, 128k, 256k, 512k, and 1024k tokens, yielding a total of 684 question-answer instances. This enables systematic analysis of model performance across an extended range of context sizes, including the challenging 1M token regime.

Questions are designed to require multi-file reasoning, understanding of class hierarchies and inheritance patterns, exception propagation across method boundaries, and semantic comprehension of API contracts. The dataset tests both shallow pattern matching and deep compositional reasoning over large Java codebases.

As with the other datasets, each question is evaluated under two conditions: multiple-choice setting, and open-ended generation without options.

\section{Results}
\label{sec:results}
We evaluate a diverse set of models, including GPT-4o~\cite{openai2024gpt4o}, 
Gemini~\cite{gemini2024}, Claude~\cite{anthropic2024claude}, 
Llama~\cite{touvron2023llama3}, Mistral~\cite{jiang2023mistral}, 
Qwen~\cite{qwen2023}, and Granite~\cite{granite2024}.

\subsection{Python}
\label{sec:results_python}

We evaluated the models on subset of LongCodeBench\cite{rando2025longcodebench} across context lengths ranging from 32k to 512k tokens. Table~\ref{tab:python_results} presents accuracy results under both the ``With Options'' (shuffled) and ``Without Options'' conditions.

\paragraph{Observations.}
Several consistent patterns emerge from the Python results. First, all models exhibit a substantial recognition versus generation gap: accuracy drops by 15--35 percentage points when options are removed. Claude Sonnet 4.5 achieves the highest accuracy with options (80.5\% at 32k) but drops to 42.5\% without them. Similarly, Llama-3.1-405B, despite its large parameter count, shows a dramatic 32-point drop at 32k context.

Second, performance does not scale uniformly with context length. Many models show degradation or inconsistent behaviour as context increases. For instance, Llama-3.1-405B drops from 76.3\% at 64k to 68.5\% at 128k in the multiple-choice setting, and further to 32.6\% without options. This suggests that simply extending context windows does not guarantee improved reasoning.

Third, Gemini models show relative robustness in the open-ended setting. Gemini-2.5-Pro maintains 44--54\% accuracy without options across all tested lengths, and Gemini-2.5-Flash achieves up to 54.3\% at 128k. This indicates stronger generation capabilities compared to other models, though still significantly below their multiple-choice performance.

Fourth, the results reveal that even frontier models struggle with free-form code reasoning at scale. The consistent accuracy drop suggests models rely heavily on recognizing answer patterns within provided options rather than independently reconstructing correct answers from code analysis.

\subsection{COBOL}
\label{sec:results_cobol}

We evaluated models on two COBOL datasets: OPPSCAL (public Medicare Pricer code) and an internal IBM dataset. Results are presented in Tables~\ref{tab:oppscal_results} and~\ref{tab:internal_results}.

\paragraph{Observations.}
The COBOL results reveal several striking patterns that differ from the Python findings.

First, models achieve exceptionally high accuracy on OPPSCAL with options—multiple models exceed 98\% accuracy at 32k and 128k contexts. This suggests that OPPSCAL questions may involve more explicit code patterns or that COBOL's procedural structure provides clearer local cues. However, this high performance completely collapses without options: accuracy drops by 30--60 percentage points across models. Gemini-2.5-Pro drops from 98.4\% to 62.3\% at 32k, and further to 32.0\% at 512k without options.

Second, the internal IBM dataset proves more challenging, with ``With Options'' accuracy ranging from 73--90\%. This suggests greater complexity or ambiguity in enterprise transaction processing logic compared to the Pricer codebase. Notably, Gemini-2.5-Pro and Gemini-2.5-Flash show consistent improvement from 32k to 512k contexts, reaching over 90\% accuracy, indicating effective utilization of long-context information.

Third, the generation gap is more severe in COBOL than Python. Even the best-performing models struggle to generate correct answers without multiple-choice scaffolding. This may reflect limited COBOL training data in pretraining corpora, making it harder for models to internalize legacy language idioms and generate syntactically correct COBOL-specific terminology.

Fourth, Gemini models show the strongest COBOL reasoning, particularly in longer contexts. Gemini-2.5-Pro maintains above 90\% accuracy with options at 256k and 512k on the internal dataset, and retains around 50\% accuracy even without options—substantially better than GPT-4o's 38.5--54.7\% and Claude's 46.7--53.3\%.

Overall, the COBOL results underscore that legacy code understanding remains fragile: models can exploit multiple-choice cues effectively but lack robust generation capabilities for COBOL-specific logic, revealing a critical gap for enterprise software maintenance applications.

\subsection{Java}
\label{sec:results_java}
We evaluate models on LongCodeBenchQA-Java across context windows ranging from 32k to 1024k tokens. Results are summarized in Table~\ref{tab:results} and Figure~\ref{fig:model-performance-wide}.

\paragraph{Observations.}

The Java results reveal several consistent patterns that complement the Python and COBOL findings.

First, models achieve strong performance in the multiple-choice setting, with several frontier models exceeding 85\% accuracy at smaller context lengths. Llama-3.3-70B and GPT-4o reach 89.5\% accuracy at 32k and 64k respectively, while Mistral-Large-675B achieves 86.8\% at 64k. However, performance does not consistently improve with longer contexts. Notably, Llama-3.3-70B drops from 85.1\% at 64k to 65.8\% at 128k, and Mistral-Small-24B drops from 82.5\% to 67.5\%, indicating that larger contexts may introduce distractors or exceed models' effective retrieval capacity.

Second, models with explicit long-context optimization demonstrate clear advantages. Gemini-2.5-Flash achieves the strongest long-context results, reaching 86.0\% at 128k and 87.7\% at 256k, while maintaining stable performance up to 1024k tokens. Gemini-2.5-Pro shows similarly stable behavior, consistently achieving 78--81\% accuracy across the full context range.

Third, the recognition versus generation gap is noticeably smaller than in the Python and COBOL datasets. Several frontier models achieve strong open-ended performance, with Mistral-Large-675B reaching 77.2\% at 64k and Gemini-2.5-Flash reaching 78.1\% at 256k. GPT-4o also maintains high open-ended accuracy (77.2--78.1\%) across supported contexts. This suggests that modern models are increasingly capable of precise free-form reasoning over Java codebases.

Fourth, mid-sized models such as Granite-4.0-8B and Qwen3-30B-Thinking show consistent but lower performance across supported context lengths, while larger frontier models dominate performance in both evaluation settings.

Overall, the Java results reinforce a key trend across languages: long-context support alone does not guarantee improved reasoning. Instead, architectural and training differences strongly influence how effectively models utilize extended repository context.



\section{Needle-In-A-Haystack Experiments}
We evaluate standard multiple-choice QA and a long-context retrieval sensitivity test modelled after \textit{Needle-In-A-Haystack}, where  information is planted at the \textbf{Start}, \textbf{Middle} and \textbf{End} of the context.

To rigorously evaluate the impact of token position on the model's retrieval capabilities within extended context windows, we adopt a modified "Needle in the Haystack" evaluation setting. We formalize the context window $\mathcal{C}$ as a sequence of tokens $\mathcal{C} = \{t_1, t_2, \dots, t_L\}$, where $L$ represents the maximum context length. The experiment is stratified into two distinct settings to decouple pure recall capability from noise robustness.

\begin{table*}[ht]
    \centering
    \resizebox{\textwidth}{!}{%
    \begin{tabular}{llccccccccc}
        \toprule
        \multirow{2}{*}{\textbf{Model}} & \multirow{2}{*}{\textbf{Method}} & \multicolumn{3}{c}{\textbf{32k}} & \multicolumn{3}{c}{\textbf{64k}} & \multicolumn{3}{c}{\textbf{128k}} \\
        \cmidrule(lr){3-5} \cmidrule(lr){6-8} \cmidrule(lr){9-11}
         & & \textbf{Start} & \textbf{Middle} & \textbf{End} & \textbf{Start} & \textbf{Middle} & \textbf{End} & \textbf{Start} & \textbf{Middle} & \textbf{End} \\
        \midrule
        
        \multicolumn{11}{c}{\textbf{Language: Java}} \\
        \midrule
        
        \multirow{2}{*}{Llama-3.1-405B} 
          & With Options & 87.72\% & 85.96\% & 85.96\% & 83.33\% & 85.09\% & 85.09\% & 73.68\% & 79.82\% & 83.33\% \\
          & Without Options & 62.28\% & 62.28\% & 65.79\% & 60.53\% & 62.28\% & 63.16\% & 52.63\% & 58.77\% & 64.04\% \\
        \midrule
        
        \multirow{2}{*}{Granite-4-8B} 
          & With Options & 77.19\% & 82.46\% & 84.21\% & 69.30\% & 72.81\% & 79.82\% & 71.05\% & 76.32\% & 78.07\% \\
          & Without Options & 57.02\% & 58.77\% & 64.91\% & 42.98\% & 54.39\% & 56.14\% & 39.47\% & 53.51\% & 64.04\% \\
        \midrule
        
        \multirow{2}{*}{Qwen3-30B-thinking} 
          & With Options & 73.68\% & 71.93\% & 72.81\% & 70.18\% & 72.81\% & 75.44\% & 70.18\% & 71.05\% & 77.19\% \\
          & Without Options & 65.79\% & 63.16\% & 67.54\% & 64.04\% & 58.77\% & 64.04\% & 62.28\% & 63.16\% & 63.16\% \\
        \midrule
        
        \multirow{2}{*}{Mistral-Small} 
          & With Options & 52.63\% & 57.02\% & 55.26\% & 66.67\% & 77.19\% & 70.18\% & - & - & - \\
          & Without Options & 57.14\% & 60.00\% & 54.29\% & 60.00\% & 57.14\% & 58.57\% & - & - & - \\
        \midrule
        
        \multirow{2}{*}{Mistral-Medium} 
          & With Options & 83.33\% & 78.95\% & 84.21\% & 82.46\% & 84.21\% & 85.96\% & - & - & - \\
          & Without Options & 64.04\% & 64.91\% & 68.42\% & 64.91\% & 66.67\% & 69.30\% & - & - & - \\

        \midrule
        \multicolumn{11}{c}{\textbf{Language: COBOL}} \\
        \midrule

        \multirow{2}{*}{Llama-3.1-405B} 
          & With Options & 84.62\% & 80.77\% & 73.08\% & 75.56\% & 82.22\% & 77.78\% & 70.67\% & 76.00\% & 84.00\% \\
          & Without Options & 65.38\% & 76.92\% & 73.08\% & 57.78\% & 53.33\% & 60.00\% & 40.00\% & 46.67\% & 60.00\% \\

        \midrule

        \multirow{2}{*}{Granite-4-8B} 
          & With Options & 76.92\% & 73.08\% & 76.92\% & 57.78\% & 62.22\% & 75.56\% & 53.33\% & 64.00\% & 81.33\% \\
          & Without Options & 53.85\% & 53.85\% & 76.92\% & 15.56\% & 35.56\% & 62.22\% & 20.00\% & 32.00\% & 69.33\% \\
          
        \midrule

        \multirow{2}{*}{Qwen3-30B-thinking} 
          & With Options & 84.62\% & 73.08\% & 76.92\% & 68.89\% & 73.33\% & 77.78\% & 80.00\% & 77.33\% & 80.00\% \\
          & Without Options & 76.92\% & 73.08\% & 76.92\% & 60.00\% & 62.22\% & 60.00\% & 69.33\% & 65.33\% & 70.67\% \\

        \midrule

        \multirow{2}{*}{Mistral-Small} 
          & With Options & 69.23\% & 50.00\% & 73.08\% & 64.44\% & 68.89\% & 68.89\% & - & - & - \\
          & Without Options & 69.23\% & 65.38\% & 76.92\% & 53.33\% & 55.56\% & 66.67\% & - & - & - \\
          
        \midrule

        \multirow{2}{*}{Mistral-Medium} 
          & With Options & 73.08\% & 80.77\% & 76.92\% & 80.00\% & 82.22\% & 82.22\% & - & - & - \\
          & Without Options & 61.54\% & 61.54\% & 73.08\% & 60.00\% & 48.89\% & 57.78\% & - & - & - \\

        \bottomrule
    \end{tabular}%
  }
        \caption{This table discusses the performance of various open-source models on the Relevant Needle Setting for Java and COBOL, where  relevant information is planted at the Start, Middle and End of the context.}
        \label{tab:Relevant_needle}  

\end{table*}

\subsection{Setting I: Relevant Needle and Positional Recall}
In the first configuration, we aim to investigate the \textit{primacy} and \textit{recency} biases inherent in Transformer-based attention mechanisms. Let $n_{rel}$ denote the specific code snippet (the "needle") required to correctly answer the query $q$. We construct a composite context $\mathcal{C}'$ by inserting $n_{rel}$ into a corpus of unrelated code ("haystack") $\mathcal{H}$ at a normalized depth $d \in [0, 1]$, where $d=0$ denotes the Start (prefix) and $d=1$ denotes the End (suffix) of the context window.

The objective is to measure the retrieval accuracy $Acc(d)$ as a function of injection depth in three different setting \textit{Start ($d=0$), Middle ($d=0.5$) and End ($d=1$)}. 
In our context, we evaluate if the generation of the correct answer $a$ fluctuates based on the position of the evidence.

Based on our experiments in Table \ref{tab:Relevant_needle}, we observe a consistent and significant performance delta between the \textit{With Options} (discriminative) and \textit{Without Options} (generative) settings. Models retain the ability to attend to the relevant information, evidenced by high scores in the multiple-choice setting, but struggle to reproduce it verbatim in open-ended generation. For instance, in Java at 128k tokens, \textbf{Llama-3.1-405B} drops from $\approx 79\%$ (With Options) to $\approx 58\%$ (Without Options). This indicates that while the attention mechanism remains functional, generation coherence degrades significantly at scale. Further, with option setting, provide the model an opportunity to match the verbatim given in the option directly with the code context, which is easier compared to open-ended generation that requires a deep understanding of the query semantics, and its alignment to the code. This indicates that while the model successfully attends to the relevant needle, it struggles to formulate the exact verbatim extraction or reasoning required for open-ended generation in long-context scenarios.

Contrary to the standard ``Lost in the Middle'' phenomenon, we note that most models exhibit recency bias, particularly in COBOL. We observe that the performance of the 'end` setting is better in comparison to the other settings of start and middle while perfromance measuure even following a steep upward trend from Start to End. For example, Granite-4-8B in without option setting shows  $15.56\%$ (for Start), $35.56\%$ (for Middle) and  $62.22\%$ (for End).

While Java performance remains relatively stable across 32k to 128k for frontier models, COBOL exposes fragility in context scaling. \textbf{Llama-3.1-405B} holds up well in Java 128k ($\approx 60\%$ generative accuracy), but in COBOL 128k, its generative accuracy at the Start position crashes to $40.00\%$. This implies that the ``effective context length'' is language-dependent; reliable retrieval windows shrink significantly for legacy or lower-resource languages. 

Overall, the results indicate that while retrieval-augmented generation (With Options) remains better compared to open-ended generation (Without Options), which suffers from severe position-dependent degradation. This is most acute in the ``Lost at the Start'' phenomenon, suggesting a failure in long-range dependency modelling for out-of-domain syntaxes.

\begin{table*}[ht]

    \resizebox{\textwidth}{!}{%
    \begin{tabular}{llccccccccc}
        \toprule
        \multirow{2}{*}{\textbf{Model}} & \multirow{2}{*}{\textbf{Method}} & \multicolumn{3}{c}{\textbf{32k}} & \multicolumn{3}{c}{\textbf{64k}} & \multicolumn{3}{c}{\textbf{128k}} \\
        \cmidrule(lr){3-5} \cmidrule(lr){6-8} \cmidrule(lr){9-11}
         & & \textbf{Start} & \textbf{Middle} & \textbf{End} & \textbf{Start} & \textbf{Middle} & \textbf{End} & \textbf{Start} & \textbf{Middle} & \textbf{End} \\
        \midrule
        
        \multicolumn{11}{c}{\textbf{Language: Java}} \\
        \midrule
        
        \multirow{2}{*}{Llama-3.1-405B} 
          & With Options & 86.84\% & 86.84\% & 86.84\% & 84.21\% & 80.70\% & 78.07\% & 76.32\% & 78.07\% & 77.19\% \\
          & Without Options & 66.67\% & 64.04\% & 63.16\% & 59.65\% & 59.65\% & 57.89\% & 54.39\% & 50.00\% & 49.12\% \\
        \midrule
        
        \multirow{2}{*}{Granite-4-8B} 
          & With Options & 75.44\% & 71.05\% & 75.44\% & 70.18\% & 69.30\% & 74.56\% & 71.93\% & 71.05\% & 72.81\% \\
          & Without Options & 59.65\% & 56.14\% & 57.89\% & 45.61\% & 46.49\% & 51.33\% & 47.37\% & 42.98\% & 39.47\% \\
        \midrule
        
        \multirow{2}{*}{Qwen3-30B-thinking} 
          & With Options & 71.05\% & 75.44\% & 73.68\% & 67.54\% & 71.93\% & 73.68\% & 70.18\% & 71.93\% & 71.93\% \\
          & Without Options & 62.28\% & 62.28\% & 59.65\% & 58.77\% & 63.16\% & 62.28\% & 64.04\% & 66.67\% & 61.40\% \\
        \midrule
        
        \multirow{2}{*}{Mistral-Small} 
          & With Options & 59.65\% & 59.65\% & 62.28\% & 72.81\% & 71.05\% & 75.44\% & - & - & - \\
          & Without Options & 61.40\% & 58.77\% & 57.02\% & 60.53\% & 56.14\% & 53.51\% & - & - & - \\
        \midrule
        
        \multirow{2}{*}{Mistral-Medium} 
          & With Options & 82.46\% & 78.95\% & 81.58\% & 79.82\% & 81.58\% & 82.46\% & - & - & - \\
          & Without Options & 68.42\% & 67.54\% & 66.67\% & 64.04\% & 63.16\% & 65.79\% & - & - & - \\

        \midrule
        \multicolumn{11}{c}{\textbf{Language: COBOL}} \\
        \midrule

        \multirow{2}{*}{Llama-3.1-405B} 
          & With Options & 88.46\% & 80.77\% & 84.62\% & 77.78\% & 73.33\% & 71.11\% & 78.67\% & 80.00\% & 78.67\% \\
          & Without Options & 69.23\% & 65.38\% & 69.23\% & 60.00\% & 60.00\% & 64.44\% & 48.00\% & 56.00\% & 54.67\% \\
        \midrule

        \multirow{2}{*}{Granite-4-8B} 
          & With Options & 80.77\% & 80.77\% & 76.92\% & 60.00\% & 57.78\% & 60.00\% & 62.67\% & 61.33\% & 61.33\% \\
          & Without Options & 50.00\% & 53.85\% & 53.85\% & 37.78\% & 28.89\% & 26.67\% & 34.67\% & 32.00\% & 32.00\% \\
          
        \midrule
        
        \multirow{2}{*}{Qwen3-30B-thinking} 
          & With Options & 76.92\% & 80.77\% & 73.08\% & 71.11\% & 75.56\% & 75.56\% & 74.67\% & 78.67\% & 74.67\% \\
          & Without Options & 76.92\% & 73.08\% & 69.23\% & 62.22\% & 60.00\% & 66.67\% & 65.33\% & 68.00\% & 57.33\% \\
          \midrule
        
        \multirow{2}{*}{Mistral-Small} 
          & With Options & 69.23\% & 65.38\% & 57.69\% & 64.44\% & 62.22\% & 62.22\% & - & - & - \\
          & Without Options & 57.69\% & 61.54\% & 65.38\% & 53.33\% & 57.78\% & 55.56\% & - & - & - \\

        \midrule
        
        \multirow{2}{*}{Mistral-Medium} 
          & With Options & 69.23\% & 73.08\% & 84.62\% & 82.22\% & 75.56\% & 71.11\% & - & - & - \\
          & Without Options & 53.85\% & 65.38\% & 61.54\% & 51.11\% & 60.00\% & 53.33\% & - & - & - \\

        \bottomrule
    \end{tabular}%
    }
        \centering
    \caption{This table discusses the performance of various open-source models on the Irrelevant Needle Setting for Java and COBOL, where irrelevant information is planted at the Start, Middle and End of the context.}
    \label{tab:irrelevant_needle}

\end{table*}

\subsection{Setting II: Irrelevant Needle and Distractor Robustness}
The second configuration evaluates the model's robustness against \textit{distractor injection}. Here, we introduce an irrelevant code snippet, denoted as $n_{irr}$, which does not share any similarities with the correct answer and contains no information relevant to answering $q$.

We systematically manipulate the position $d_{irr}$ of this noisy needle to \textit{Start ($d=0$), Middle ($d=0.5$) and End ($d=1$)} while keeping the relevant information $n_{rel}$ fixed. This setting determines if the model's attention mechanism can effectively filter high-rank noise located at privileged positions (e.g., $d_{irr}=1$).

Based on our experiments in Table \ref{tab:irrelevant_needle}, we observe that across all models and context lengths (32k, 64k, and 128k), the \textit{"With Options"} configuration consistently outperforms the \textit{"Without Options"} setup, similar to the relevant needle section. For instance, Llama-3.1-405B in Java at 128k shows a performance gap of over $20\%$ between the two methods ($76.32\%$ vs $54.39\%$), suggesting that providing a constrained search space significantly eases the task.

Performance trends on both languages are similar for larger models; however, for smaller models,  COBOL generally lags behind Java, particularly for smaller models. Interestingly, Qwen3-30B-thinking demonstrates high resilience in COBOL, maintaining relatively stable performance across 32k to 128k (ranging from $71.05\%$ to $78.67\%$ in the \textit{With Options} setting), outperforming even larger models in specific long-context scenarios. 

Most models exhibit relative stability regardless of whether the distractor is placed at the \textit{Start, Middle}, or \textit{End}. However, larger context windows (128k) begin to expose slight vulnerabilities, where performance variations increase and a general drop is observed when the irrelevant needle is placed at the end, again showing evidence for recency bias aligning with the observations of relevant needle setting.

\section{Conclusion}
\label{sec:conclusion}
Our study reveals that long-context code QA remains far from solved. Despite million-token context windows, LLMs remain vulnerable to perturbations that require genuine reasoning rather than pattern recognition. Future work should prioritise mitigations for positional sensitivity and improved compositional reasoning over large codebases.

\bibliography{custom}

\appendix


\end{document}